\documentclass[sigconf]{acmart}

\usepackage{subfigure}
\usepackage{multirow}
\usepackage{float}
\usepackage{fancyhdr}
\AtBeginDocument{%
  \providecommand\BibTeX{{%
    \normalfont B\kern-0.5em{\scshape i\kern-0.25em b}\kern-0.8em\TeX}}}

\setcopyright{acmcopyright}
\copyrightyear{2021}
\acmYear{2021}
\setcopyright{acmcopyright}\acmConference[MM '21]{Proceedings of the 29th ACM
International Conference on Multimedia}{October 20--24, 2021}{Virtual Event, China}
\acmBooktitle{Proceedings of the 29th ACM International Conference on Multimedia
(MM '21), October 20--24, 2021, Virtual Event, China}
\acmPrice{15.00}
\acmDOI{10.1145/3474085.3475437}
\acmISBN{978-1-4503-8651-7/21/10}

\acmSubmissionID{1508}

\begin{document}
\fancyhead{}

\title{Multi-Singer: Fast Multi-Singer Singing Voice Vocoder With A Large-Scale Corpus}

\author{Rongjie Huang, Feiyang Chen, Yi Ren, Jinglin Liu, Chenye Cui, Zhou Zhao}
\authornote{Corresponding author.}
\email{{rongjiehuang, chenfeiyang, rayeren, jinglinliu, chenyecui, zhaozhou}@zju.edu.cn}
\affiliation{%
  \institution{Zhejiang University}
  \country{}
}


\begin{abstract}
High-fidelity multi-singer singing voice synthesis is challenging for neural vocoder due to the singing voice data shortage, limited singer generalization, and large computational cost. Existing open corpora could not meet requirements for high-fidelity singing voice synthesis because of the scale and quality weaknesses. Previous vocoders have difficulty in multi-singer modeling, and a distinct degradation emerges when conducting unseen singer singing voice generation. To accelerate singing voice researches in the community, we release a large-scale, multi-singer Chinese singing voice dataset OpenSinger. 
To tackle the difficulty in unseen singer modeling, we propose Multi-Singer, a fast multi-singer vocoder with generative adversarial networks. Specifically, 1) Multi-Singer uses a multi-band generator to speed up both training and inference procedure. 2) to capture and rebuild singer identity from the acoustic feature (i.e., mel-spectrogram), Multi-Singer adopts a singer conditional discriminator and conditional adversarial training objective. 3) to supervise the reconstruction of singer identity in the spectrum envelopes in frequency domain, we propose an auxiliary singer perceptual loss. The joint training approach effectively works in GANs for multi-singer voices modeling. Experimental results verify the effectiveness of OpenSinger and show that Multi-Singer improves unseen singer singing voices modeling in both speed and quality over previous methods. The further experiment proves that combined with FastSpeech 2 as the acoustic model, Multi-Singer achieves strong robustness in the multi-singer singing voice synthesis pipeline. Samples are available at \url{https://Multi-Singer.github.io/}

\end{abstract}




\maketitle

\section{Introduction}

Singing voice synthesis (SVS) aims to synthesize high-quality and expressive singing voices based on musical score information. Singing voice synthesis (SVS) systems~\cite{gu2020bytesing,chen2020hifisinger,liu2021diffsinger} take music score and lyric information as input to generate singing voices, and these systems have been widely deployed in music softwares, music boxes, and so on. SVS systems could generate singing voices with comparable quality to reference songs, which attract widespread research interest.

Following the essential components similar to TTS systems, SVS systems generally adopt an acoustic model~\cite{lu2020xiaoicesing,Korean} to convert the musical scores into acoustic features, and a vocoder~\cite{kumar2019melgan,chen2020hifisinger} to generate audio waveform from acoustic features. Neural vocoders can synthesize natural-sounding speech, which generally determines the upper bound of generated sound quality. In this paper, we concentrate on waveform modeling in vocoder. 

Unlike traditional TTS~\cite{ren2019fastspeech,chen2020multispeech,ren2020simulspeech,ren2020study,zhang2021wsrglow}, there are several challenges to build a multi-singer SVS system: 1) Open source and high-quality singing voice data. Unlike speech, high-quality singing voices are commonly produced by professional singers. Because of the high cost of recording and labeling songs, researchers hardly have access to large and open-source singing voice corpora. 2) Fast audio synthesis against limited computation resource. For application deployment, generation speed and computational cost need further consideration. 3) Multi-singer modeling. Timbres could be wildly different among groups while singing voices vary from expression and style. When applying SVS systems for unseen singer modeling, there comes distinct degradation in synthetic singing voice quality. 

In the past few years, researchers work to address the challenges above in singing voices modeling, while some problems emerge: 1) SVS systems like DeepSinger~\cite{DeepSinger} mine data from the web, but processed data with noise still could not meet requirements for high-fidelity SVS synthesis. Further, although several singing voice datasets such as MIR-1K dataset~\cite{hsu2009improvement} and JukeBox~\cite{chowdhury2020jukebox} have been released for research purposes, but the corpora are not so large as expected for multiple tasks. 2) Several parallel generation methods~\cite{yang2020multiband,yu2019durian} have been proposed to speed up waveforms synthesis. However, existing multi-band architectures do not consider characteristic differences among frequency bands, so a powerful frequency-adapted multi-band technique is required. 3) Researchers investigate ways to generate high-quality waveforms during multi-singer modeling. Previous multi-speaker data training strategy~\cite{cooper2020pretraining} increases model generalization. Unfortunately, without explicitly adopting architecture for singer identity reconstruction, vocoders would be data-hungry, and generalization restriction still comes. Giving additional information during generation would be another strategy. Speaker Conditional WaveRNN~\cite{paul2020speaker} takes speaker embeddings as additional input, while extra embedding would be sometimes hard-earned during inference procedure and slows down generation as well.

To accelerate SVS research, we assemble an open-source, large-scale, and multi-singer singing voice corpus OpenSinger. To the best of our knowledge, OpenSinger is the first open-source Chinese singing voice dataset. We have attached part of OpenSinger to the supplementary materials, and we will release the entire dataset after paper publication. 
To overcome afore problems in this paper, we propose Multi-Singer, which achieves computational efficiency and keeps powerful capability for multi-singer singing voices modeling. To be more specific, 1) we introduce a novel multi-band generator, which speeds up singing voice generation and improves the audio quality of different frequency bands. 2) Then we introduce the singer conditional discriminator with conditional loss function for high quality and similarity multi-singer singing voice generation. 3) To further reconstruct singer representations in the frequency domain, we propose an auxiliary singer perceptual loss based on embeddings extracted from a pre-trained speaker encoder. The proposed training method effectively works in GANs for multi-singer singing voice modeling. 

Experimental results show that Multi-Singer can generate high-fidelity multi-singer singing voices and achieve the best mean opinion score (MOS) among existing neural vocoders. Combined with FastSpeech 2 as an acoustic model, Multi-Singer shows strong robustness in singing voice synthesis systems. Multi-Singer is substantially faster than most neural vocoders, and it samples 125 times faster than real-time on single NVIDIA 2080Ti GPU with comparable quality to an autoregressive counterpart.

\section{Related Works}
In this section, we overview existing datasets, provide a singing voice synthesis background and briefly review several variations of vocoders.

\subsection{Dataset}
Training TTS and SVS systems both require a significant amount of annotated data\cite{cui2021emovie,hsu2009improvement,duan2013nus}. The rapid increase in the amount of multimedia content on the Internet in recent years makes data much more important. Researchers have released speech and singing voice corpora, varying from languages, speakers, and so on. 

\subsubsection{Speech}
\
\newline
\indent For speech synthesis, various datasets are available for different tasks. LSSED~\cite{fan2021lssed} is a challenging large-scale English speech emotion dataset, which has data collected from 820 subjects to simulate real-world distribution. AISHELL-3~\cite{shi2020aishell} contains roughly 85 hours of emotion-neutral recordings spoken by 218 native Chinese speakers, which could be applied for multi-speaker speech synthesis. Data could be collected with different methods, including mining from webs automatically, recording manually, and so on. The read English speech LibriSpeech corpus~\cite{panayotov2015librispeech} is derived from audiobooks. Japanese multi-speaker singing-voice corpus JVS-MuSiC~\cite{tamaru2020jvs} is produced in a recording studio, and the recordings were controlled by a professional sound director.

\subsubsection{Singing voices} 
\
\newline
\indent Singing voice data differs greatly from speech. Our preliminary research concludes that the main differences between singing voices and speech lie in phoneme duration and pitch (i.e., fundamental frequency), which we would discuss in appendix A in the supplementary materials. Unlike TTS with sufficient transcribed data, SVS suffers from data shortage due to its high recording and annotation cost and stricter copyright issues in the music domain. Limited singing datasets of different sizes and annotated contents are available for research purposes, and here we introduce a few singing voice datasets for comparison in Table~\ref{matrix:data}. 

The MIR-1K dataset~\cite{hsu2009improvement} establish the first comprehensive and publicly available dataset for singing voice separation, which does not contain segmentation on the word level. The proposal of multi-singer NUS-48E corpus~\cite{duan2013nus} is an ongoing effort toward a comprehensive, well-annotated dataset for singing voice related research. But NUS-48E dataset includes 48 songs with reasonably balanced phoneme distribution, which is not large enough for SVS systems training. JukeBox~\cite{chowdhury2020jukebox} contains 467 hours of singing audio data sampled at 16 kHz, which has been downloaded from the Internet Archive (IA) with a wide variety of acoustic environments and recording apparatus. Hence, it is not suitable for high-quality singing voice synthesis. 

To conclude, the above data could not meet our requirements for singing voice synthesis research in terms of quality and quantity. Here in this paper, we propose an open-source, large-scale, and multi-singer singing voice corpus OpenSinger.

\subsection{Singing voice synthesis}
Singing voice synthesis (SVS) is a generative task that produces acoustic waveforms of singing given lyrics and music score input. A typical singing voice synthesis system consists of an acoustic model to convert musical scores into acoustic features and a vocoder to generate audio waveforms from acoustic features. Previous works have conducted studies on SVS from multiple aspects. DeepSinger~\cite{DeepSinger} is a multi-lingual SVS system built from scratch using singing training data mined from music websites. Choi at all~\cite{choi2020korean} build a Korean singing voice synthesis system using an auto-regressive algorithm that generates spectrogram with the boundary equilibrium GAN objective. Chen at all~\cite{chen2020hifisinger} introduce multi-scale adversarial training in both the acoustic model and vocoder to improve singing modeling. As the papers say, these previous SVS systems could generate natural singing voices. However, because vocoders in such SVS systems are not designed towards multiple singers, there would be a distinct degradation in quality when generating unseen singers' voices.

\begin{table}[]
  \centering
  \small
  \vspace{-2mm}
  \begin{tabular}{ccc}
  \toprule
  Name           & Task                     & Language               \\
  \midrule
  NUS-48E corpus & singing voice research   & English                \\
  MIR-1K dataset & singing voice separation & Chinese                \\
  JukeBox        & singer recognition       & 18 different languages \\ 
  OpenSinger       & singing voice synthesis  & Chinese                \\
  \bottomrule
  \end{tabular}
  \caption{Usage and language of datasets.}
  \label{matrix:data}
  \end{table}

\subsection{Vocoder}
With the powerful model assumption and the solid theoretical ground, autoregressive vocoders have dominated singing voice modeling for a long time. WaveRNN~\cite{kalchbrenner2018efficient} is an autoregressive model regularly adopted to synthesize waveform in SVS systems. Non-autoregressive vocoder Parallel WaveNet~\cite{oord2018parallel} provides a fast waveform generation method based on a teacher-student framework with probability density distillation.
Besides, WaveGlow~\cite{prenger2019waveglow} is another non-autoregressive vocoder, which consumes an enormous computation cost during training. More recently, researchers propose diffusion-based models WaveGrad~\cite{chen2020wavegrad} and DiffWave~\cite{kong2020diffwave} for waveform generation, which are built on prior work of score matching and diffusion probabilistic models.

To avoid the sample-by-sample causal inference or the use of teacher models, researchers have adopted the idea of the generative adversarial network (GAN) to train neural vocoders. MelGAN~\cite{kumar2019melgan} is a light non-autoregressive vocoder pioneering based on a generative adversarial network, which is free from distillation. HiFi-GAN~\cite{kong2020hifigan} consists of small sub-discriminators obtaining specific periodic parts of raw waveforms, achieving higher computational efficiency and sample quality. VocGAN~\cite{yang2020vocgan} applies the joint conditional and unconditional objective, which is inspired by successful results in high-resolution image synthesis. Although these vocoders could be applied in SVS systems, distinct degradations occur when generalizing those systems to unseen singers.

\section{Construction of OpenSinger}
OpenSinger contains pop songs collected from 93 singers, and singing utterances are saved in wav format, sampled at 24 kHz, and quantized by 16 bits. OpenSinger consists of 50 hours of singing voices recorded in a professional recording studio, including 30 hours from 41 females and 20 hours from 25 males apart from the person-of-interest (POI). Figures in appendix A in the supplementary materials summarize the distribution of pitch, sentence-level duration, and phoneme-level duration of utterances. The major features of OpenSinger include:
\begin{itemize}
\item Open source. Lack of data could obstruct the construction of SVS systems, so we release our corpus to accelerate research in the community.
\item Large scale. Data-hungry singing voice systems need a significant amount of data in the training process. To our best knowledge, OpenSinger is the most extensive Chinese multi-singer singing voice corpus.
\item High quality. Similar to text-to-speech, high-quality audios without noise or background sound are essential for high-fidelity singing voice synthesis. Professional singers and studios both ensure high-quality utterances in OpenSinger.
\end{itemize}

\subsection{Data Collection}

\quad \textbf{Collection Procedure} In the data collection procedure, we select Chinese traditional and pop songs, and organize a group of 93 professional singers to record 80 hours of singing voices. The recording takes place in a private recording studio. The songs are saved in wav format, sampled at 24 kHz, and quantized by 16 bits.

\textbf{Data labeling} A professional annotation team annotates the utterances in each song. Each utterance is annotated with the name of the song, singer, and reference text, which is the official Latinized notation for marking Chinese pronunciations. Further, We use open source tools pypinyin\footnote{\url{https://github.com/mozillazg/python-pinyin}} to convert Chinese lyrics into phonemes.

\subsection{Processing}
After data collection, the singing voice corpus still could not meet high-quality singing voice generation requirements for the following aspects: 1) Most vocoders need data in the form of relatively short utterances usually up to a few seconds in length, each with corresponding text during training procedures. 2) Pauses and resulting silences commonly remain in raw songs, which would hurt adversarial stability and bring unnecessary calculations. Our processing procedure consists of several stages to fix the issues we listed above for high-quality singing voice generation, including segmentation, silence trimming, and alignment.

\textbf{Silence trimming} After data division, there still usually be some long-term silences in singing voices. Voice Activation Detection (VAD) has been used to remove silent segments, and thus we detect and discard the non-vocal segments using VAD. During silence trimming, each utterance has been manually verified to discard audio samples that do not contain singing vocals every 100ms. Cutting these silence segments could significantly shorten the audios as exceedingly speeds up the alignment formation.

\textbf{Segmentation}  Obviously, singing voices with lengthy audio clips are not suitable for memory-limited GPU computation, so we fragment them into many small files. Following the Lyrics-to-Singing alignment in~\cite{DeepSinger}, 
we split the whole song into aligned lyrics and audio. We segment an audio by the frames that are aligned to the separation marks in raw lyrics, making sure each processed sentence's duration is limited to 0-11 seconds. 

\textbf{Alignment} For more precise alignment, we adopt Montreal Forced Aligner\footnote{\url{https://github.com/MontrealCorpusTools/Montreal-Forced-Aligner}} and take an orthographic transcription of an audio file and generate a time-aligned version. In Montreal Forced Aligner, an annotation for phonemes is generated by aligning the manually-labeled phone strings of the sung lyrics using training conventional Gaussian Mixture Model (GMM) – Hidden Markov Model (HMM) system. The annotated phonemes support us for broad and preliminary observations about the phoneme-level duration of the corpus.

\subsection{Statistics}
After the data collection and processing procedure, we check for audio quality and conduct statistical evaluation, including sentence-level and phoneme-level duration distribution, pitch distribution and speaker similarity.

\textbf{Sentence-Level Duration Distribution} We have segmented the songs into sentence-level singing voices during audio processing. We plot the sentence-level duration distribution of the male and female singing voices separately. As shown in appendix A in the supplementary materials, sentence durations between genders are similar with some small variations due to our manual segmentation in advance.

\textbf{Phoneme-Level Duration Distribution} We align the phoneme sequence to the singing audio frames using MFA mentioned above and visualize distribution in appendix A in the supplementary materials. Phoneme durations of male and female singing voices have a similar distribution, mainly scatters around 10 ms to 500 ms.
 
\textbf{Pitch Distribution} We extract pitch from the audio using Parselmouth\footnote{\url{https://github.com/YannickJadoul/Parselmouth}} and show statistical distribution. Obviously, due to the diversity of timbre and pronunciation habits among genders, the drawn pitch distribution figures differ greatly. Female singing voices exhibit a larger pitch range, trending toward high frequency more significantly. In other words, there are more high-frequency parts in female singing voices.

\begin{figure}[]
    \centering
    \vspace{-2mm}
    \includegraphics[width=0.4\textwidth]{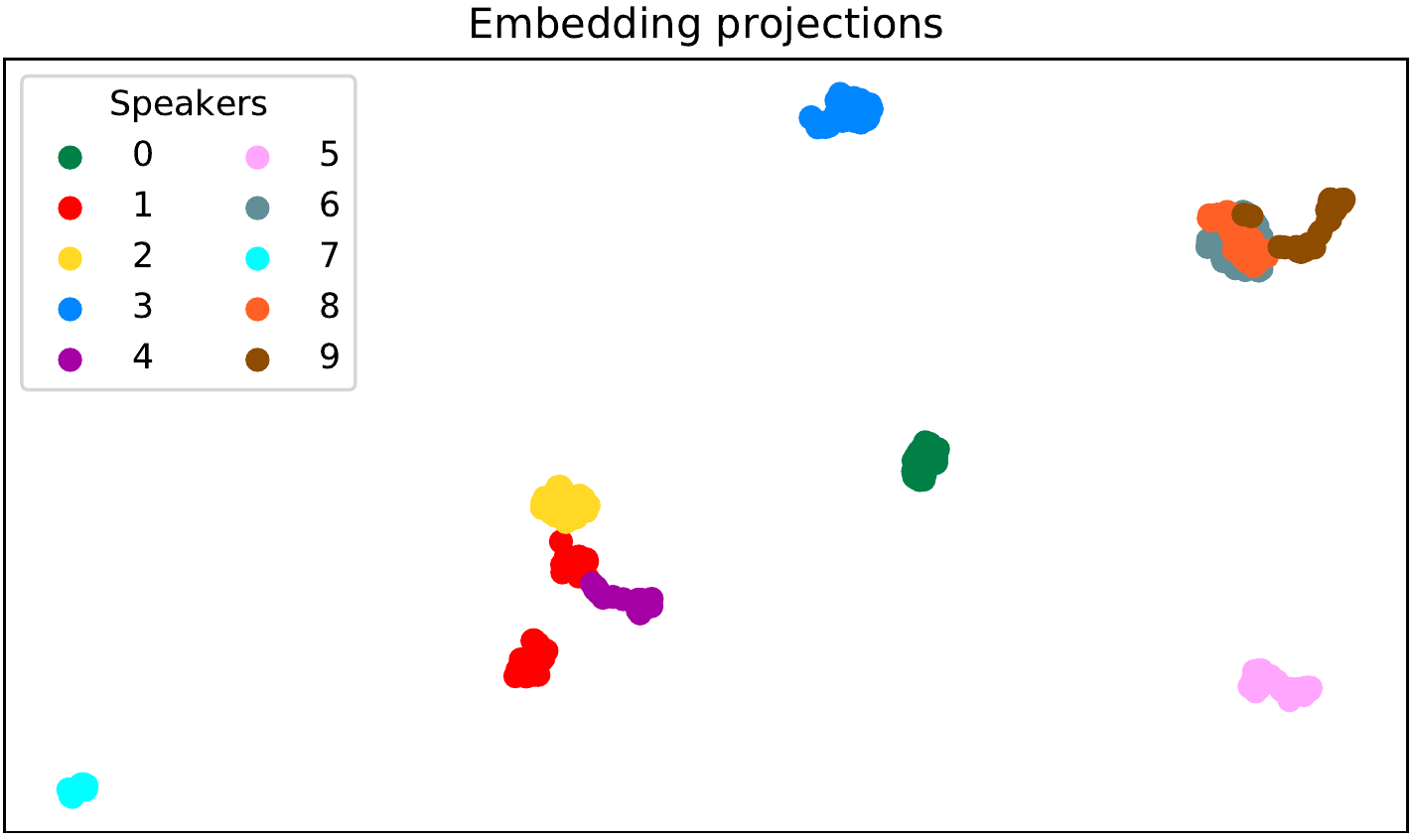}
    \caption{UMAP projection of 10 utterances for each of the 10 singers. Different colors represent different singers.}
    \label{fig:Sing_Speaker}
    \vspace{-3mm}
    \end{figure}

\textbf{Speaker Similarity} To learn about speaker identity differences among singers in OpenSinger, we visualize the speaker representations in Figure~\ref{fig:Sing_Speaker} using Resemblyzer\footnote{\url{https://github.com/resemble-ai/Resemblyzer}}. 10 utterances of 10 randomly sampled speakers are chosen, each is converted into a 256-dimensional embedding and reduced to 2-dimensional with Uniform Manifold Approximation and Projection (UMAP). It has presented a large inter-speaker distance among singers and demonstrated the singer diversity in OpenSinger.

\subsection{Speech and Singing voices}
Several characteristic differences exist between singing voice and speech generation: 1) Singing voices include richer emotional information, which varies from singing expression and style. 2) Singing voices have longer continuous pronunciations and contain more high-frequency parts. 
To verify our thinking, we conduct statistical analysis on the Chinese speech dataset CSMSC\footnote{\url{https://github.com/mozillazg/python-pinyin}} and plot the phoneme-level duration as well as pitch distribution. Because sentences are cut and segmented manually, we haven't considered sentence-level discrepancies. Table~\ref{matrix:statistics} come a few interesting phenomena: 1) Since singing voices are more various and diverse, statistic distributions of singing voices become broader with large variance. 2) Pitch and phoneme-level duration in speech corpus are lower than that in singing voices on average, demonstrating the longer continuous pronunciation and more high-frequency parts in singing voices.

To conclude, the characteristic differences between speech and singing voice we analyze above have brought additional challenges for singing voices synthesis researches, and reduce researchers' enthusiasm to assemble singing voice corpora. To our best knowledge, OpenSinger is the largest open-source, multi-singer, Chinese singing voice corpus, and we hope that the release of OpenSinger could contribute to the community.

\begin{table}[]
  \centering
  \small
  \vspace{-2mm}
  \begin{tabular}{cclcc}
  \toprule
  \multirow{2}{*}{Item} & \multicolumn{2}{c}{Pitch (Hz)}    & \multicolumn{2}{c}{Phoneme duration (ms)} \\
                        & mean   & \multicolumn{1}{c}{std} & mean                & std                 \\
  \midrule
  Singing Voice (OpenSinger)       & 280.36 & 94.56         & 186.15              & 143.92              \\
  Speech (CSMSC)               & 250.97 & 60.63            & 128.85              & 78.45                \\
  \bottomrule
  \end{tabular}
  \caption{Statistics for pitch and phoneme-level duration.}
   \vspace{-2mm}
  \label{matrix:statistics}
  \end{table}
\section{Multi-Singer}

\subsection{Motivation}
\subsubsection{Multi-band generartion}
\
\newline
\indent In recent years, state-of-the-art vocoders have significantly improved the singing voice quality and spawned SVS systems deployments. However, due to the high computational cost and time-consumed generation, real-time applications could become challenges. Researchers have adopted multi-band generation such as Multiband-WaveRNN~\cite{yu2019durian}, Multiband-MelGAN~\cite{yang2020multiband} to speed up waveform modeling. Related multi-band vocoders generate each sub-band of waveform, and then conduct bands splicing using Pseudo Quadrature Mirror Filter Bank (PQMF)~\cite{nguyen1994near}. Nevertheless, previous multi-band techniques are designed towards fast generation and lack considering the characteristic differences among frequency bands (e.g., short-period high-frequency band and long-period low-frequency band), usually resulting in the limitation of synthetic singing voice quality. This paper proposes the frequency-adaptive multi-band generation technique to adjust singing voice synthesis in both speed and quality.

\subsubsection{Multi-singer modeling}
\
\newline
\indent Sufficient singer generalization means that the vocoder could generate high-fidelity audio in various singer domains, regardless of whether the input has been encountered during training or has come from an out-of-domain singer. Researchers have investigated ways to supervise vocoders for learning singer identity in multi-singer modeling. 

\textbf{Multi-speaker data} Training on data of multiple speakers could probably improve model generalization. Using multi-speaker data, researchers~\cite{cooper2020pretraining} demonstrate better performance in speaker similarity for multi-speaker speech synthesis. However, without explicitly adopting architecture for speaker identity reconstruction, vocoders would be data-hungry and encounter generalization restriction. A distinct degradation emerges when we adapt these vocoders to the unseen speakers modeling.

\textbf{Extra embedding input} Taking embedding as additional inputs could be another way to handle singer generalization. Transfer learning from speaker verification to multi-speaker text-to-speech synthesis~\cite{jia2018transfer} takes the speaker verification network as the speaker encoder and concatenates the generated speaker embedding to each encoder time step. Speaker Conditional WaveRNN~\cite{paul2020speaker} takes speaker embeddings extracted from a pre-trained speaker verification model as additional input and exploits extra speaker information. However, extra embeddings would sometimes be hard-earned in SVS systems during the inference process. Worse still, it takes a higher computation cost and reduces inference speed intolerably. 

To maintain feasibility and improve generalization towards the unseen singers, we have better explore how to teach vocoders explicitly capture singer identity embed in the acoustic feature (i.e., mel-spectrogram) without additional computational cost during singing voice synthesis.

\subsection{Overview}
  \begin{figure*}[]
    \centering
    \vspace{-2mm}
    \includegraphics[width=0.9\textwidth]{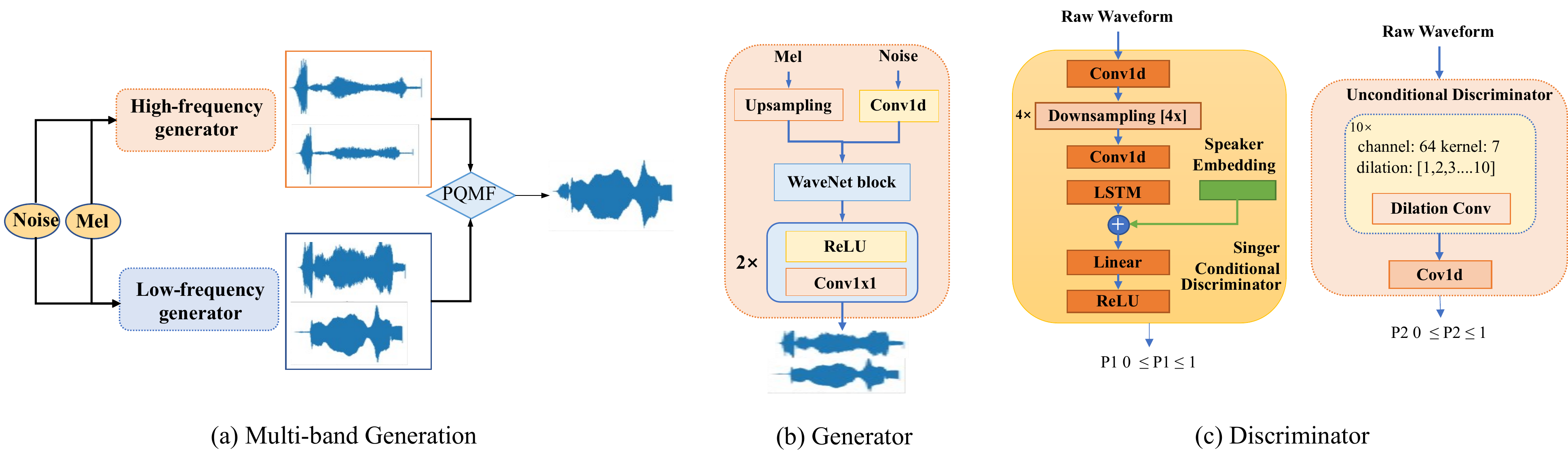}
  \vspace{-2mm}
  \caption{An architecture of Multi-Singer. (a) Multi-band generator which consists of two WaveNet-block-based generators. (b) WaveNet-block-based generator. (c) Singer Conditional and Unconditional Discriminators.}
  \label{fig:architecture} 
  \vspace{-3mm}
  \end{figure*}

Generative adversarial network based vocoder jointly trains a powerful generator G, and convolutional neural network (CNN) discriminator D, to generate time-domain waveform from the corresponding input mel-spectrogram. We have introduced a technique to improve inference speed and quality of synthetic waveform in adaptation to the unseen singer: 1) Firstly, to speed up waveforms modeling, we introduce a multi-band generator, which synthesizes sub-band signals adapted to the frequencies. 2) For multi-singer singing voice generation, Multi-Singer adopts a singer conditional discriminator to judge whether the singer identity of input voices has been properly constructed. 3) For training objective, we introduce joint adversarial training of conditional and unconditional loss and propose singer perceptual loss to penalize the generator for synthesizing singing voices with singer identification bias. The joint training method effectively works in GANs for raw audio generation.

\subsection{Generator}
To develop high-fidelity waveform generation in parallel, we introduce a multi-band generator. It is acknowledged that waveform signals' characteristics vary in different frequency bands, so we model them separately. The multi-band generator transforms the input noise drawn from a Gaussian distribution to the output waveform in parallel, and the multi-band generation process has been shown in Figure~\ref{fig:architecture}. We divide waveform into four frequency bands including two high-frequency bands and two low-frequency bands, and the waveform generation in high and low frequency bands is conducted separately using two distinct frequency-adapted models, respectively. The synthetic sub-band waveforms are merged into the final singing voices through the PQMF filter.

As shown in Figure~\ref{fig:architecture}, the generator consists of WaveNet blocks, whose architecture has been discussed in appendix E in the supplementary materials. Specifically, receptive fields and the number of WaveNet blocks vary in these two frequency-adapted models, and they are carefully designed towards different acoustic characteristics among frequency bands.  The generator is non-autoregressive and capable of adjusting singing voice synthesis in both speed and quality.

\subsection{Discriminator}

The architecture of Discriminators has been shown in Figure~\ref{fig:architecture}(c), where p1 and p2 denotes the possibility of the sample generated by conditional and unconditional discriminator, respectively. Conditional input has been applied for high-resolution image synthesis tasks and produced successful results~\cite{zhang2018stackgan++}, and conditional adversarial networks have been widely acknowledged for stable performance in image-to-image translation~\cite{isola2017image}. 
Previous vocoder studies have further demonstrated the efficiency of conditional input in discriminators. VocGAN~\cite{yang2020vocgan} introduces the hierarchically-nested JCU discriminator, which learns intermediate representations directly conditioned on the input mel-spectrogram. GAN-TTS\cite{binkowski2019high} proposes conditional DBlock, where the embedding of the linguistic features are added after the first convolution.

To better supervise the singer identity reconstruction in multi-singer singing voice generation, we adopt a singer conditional discriminator. The singer conditional discriminator judges whether the singer identity of input voices has been properly constructed, and the generator is trained to fool the singer conditional discriminator by increasing the real possibility of the generated sample.

The singer conditional discriminator(p1) first adopts a 256x downsampled block, which is performed using strided average pooling with kernel size 4 and done in 4 stages of 8x, 8x, 2x, and 2x downsampling. After passing through the downsampled block, raw waveforms are converted into 256-dimension vectors. Because singer identifications keep stable in the long-term waveforms, we feed these 256-dimension representations in LSTM layers and obtain steady identity in the final output. We conduct element-wise addition operation between high-level singer identities and reference singer embeddings, and project them to possibility with a linear layer and ReLU activation function.

The unconditional discriminator(p2) consists of ten layers of non-causal dilated 1-D convolutions. The strides are set to one and linearly increasing dilations are applied for the 1-D convolutions starting from one to eight except for the first and last layers. Channels and kernel sizes are set to 64 and 5, respectively. 

\subsection{Training Loss}
Training objectives should be carefully designed for stable training and faster convergence. In this section, we first describe \textbf{Joint adversarial conditional and unconditional loss}, and then we propose an auxiliary training loss named \textbf{Singer Perceptual Loss}. Finally, we adopt \textbf{multi-resolution STFT loss} as additional auxiliary loss and decide our final loss function.

\textbf{Joint adversarial conditional and unconditional (JCU) loss}
In contrast to conventional adversarial loss, JCU loss combines the conditional and unconditional adversarial losses as Eq.~\ref{equ:dis}. Previous work like VocGAN~\cite{yang2020vocgan} has demonstrated the efficiency of JCU loss for adversarial training. Here our adversarial conditional loss is significantly different from that in VocGAN. For multi-singer modeling, our proposed conditional loss is conducted by combating between generator and singer conditional discriminator, leading generator to better capture singer identity embed in the acoustic feature input (i.e., mel-spectrogram). The adversarial unconditional loss enhances the generator to synthesize more natural singing voices by classifying ground truth samples to 1 and the synthetic samples to 0. The JCU loss is defined in Eq.~\ref{equ:dis} and Eq.~\ref{equ:gen}.

\begin{equation}
  \begin{split}
  L_{a d v}(D ; G)=\frac{1}{2} \mathbb{E}_{x}\left[(D(x)-1)^{2}+D(y)^{2}\right]+ \\ \frac{1}{2} \mathbb{E}_{x, s}\left[(D(x, s)-1)^{2}+D_{s}(y, s)^{2}\right], 
  \label{equ:dis}
  \end{split}
  \end{equation}
 
\begin{equation}
    L_{a d v}(G ; D)=\frac{1}{2} \mathbb{E}_{y}\left[(D(y)-1)^{2}\right]+\frac{1}{2} \mathbb{E}_{y, s}\left[(D_{s}(y, s)-1)^{2}\right],
    \label{equ:gen}
\end{equation}

where $x$, $y$ and $s$ denote the ground truth and synthetic singing voices, and the singer embedding respectively. $G$ is the multi-band generator, $D$ is the unconditional discriminator, $D_s$ is the singer conditional discriminator.

\textbf{Singer Perceptual Loss}
In this work, we propose an auxiliary loss function that could improve the vocoder performance for multi-singer singing voice modeling. Perceptual loss is introduced in style reconstruction by Gratys et al.~\cite{gatys2015neural} for the first time, and many studies have taken the ideas behind perceptual loss to improve the quality of the outputs generated by a deep-learning-based model. Here we introduce a new loss objective named singer perceptual loss, which enables the generator to sense singer bias and optimize speaker similarity between ground truth and synthetic waveforms in the frequency domain during training. 

Singer perceptual loss depends on high-level hidden states extracted from the pre-trained speaker encoder, which we will discuss in appendix C in the supplementary materials. We judge mel-spectrogram in the frequency domain to calculate singer perceptual loss. It is acknowledged that the spectrum envelope of waveforms could provide singer representations, and hence envelope reconstruction in the frequency domain is also the procedure of singer identity reconstruction. Singer perceptual loss supervises singer identification reconstruction more efficiently, and it is defined as follows:

\begin{equation}
  \nonumber
  L_{spl}(x, y)= \sum_{j=1}^{L}\left(\left\|\phi_{j}(\operatorname{Mel}(x))-\phi_{j}(\operatorname{Mel}(y))\right\|_{2}\right),
\end{equation}

where $L_{spl}(x, y)$ denotes singer perceptual loss, $\|\cdot\|_{2}$ denotes the L2 norms, and let $\phi_{j}(\operatorname{Mel}(x))$ be the $j$-th layer hidden state of LSTM in the pre-trained speaker encoder $\phi$ when processing the mel-spectrogram of signal $x$. 

\textbf{Multi-resolution STFT Loss}
To further stabilize the adversarial training process, Multi-Singer adopts a multi-resolution STFT(Short Time Fourier Transform) auxiliary loss. Similar to the previous work~\cite{yamamoto2020parallel}, we define the STFT loss as follows:
\begin{equation}
  \nonumber
    L_{m_{-} s c}(x, y)=\frac{\|\operatorname{STFT}(x)-\operatorname{STFT}(y)\|_{F}}{\|\operatorname{STFT}(x)\|_{F}},
\end{equation}
\vspace{-2mm}
\begin{equation}
      \nonumber
      L_{m_{-} mag}(x, y)=\frac{1}{N}\|\log (\operatorname{STFT}(x))-\log (\operatorname{STFT}(y))\|_{1},
\end{equation}
where $\|\cdot\|_{F}$ and $\|\cdot\|_{1}$ denote the Frobenius and L1 norms; $S T F T(\cdot)$ and $N$ denote the STFT magnitude of the $m$-th STFT parameter set and the number of elements in the magnitude, respectively. $L_{m_{-}sc}$ and $L_{m_{-} mag}$ denote the spectral convergence and log STFT magnitude, respectively.

The final multi-resolution STFT loss is the sum of M losses with different analysis parameters (i.e., FFT size, window size, and hop size), which is represented as follows:
\begin{equation}
\nonumber
L_{s t f t}(x, y)=\frac{1}{M} \sum_{m=1}^{M}\left(L_{m_{-} s c}^{(m)}(x, y)+L_{m_{-} m a g}^{(m)}(x, y)\right).
\end{equation}

\textbf{Final loss}
As mentioned above, our auxiliary loss consists of the singer perceptual loss and multi-resolution STFT loss as follows:
\begin{equation}
\nonumber
  L_{a u x}(G)=\mathbb{E}_{x,y}\left[\frac{1}{2}\left(L_{spl}(x, y)+L_{s t f t}(x, y)\right)\right],
\end{equation}
where $L_{s t f t}(x, y)$ denotes the multi-resolution STFT loss, and $L_{a u x}(G)$ denotes the auxiliary loss of generator.

To conclude, our final loss function for the generator is defined as a linear combination of the auxiliary loss, adversarial loss:
\begin{equation}
\nonumber
  L_{G}=L_{a u x}(G)+\lambda L_{a d v}\left(G ; D\right),
\end{equation}
where $\lambda$ denotes the hyperparameter balancing loss terms and we set $\lambda=10$. By jointly optimizing the waveform-domain adversarial loss and auxiliary loss including singer perceptual loss and multi-resolution STFT loss, the generator can learn the distribution of the realistic speech waveform effectively.

\section{Experiment}
In this section, we first describe the experimental setup including dataset and model configurations. Then we report experimental results and conduct some analyses. 

\subsection{Experimental Setup} \label{section:Experimental Setup}
We randomly choose 340 utterances for validation and 60 utterances from 6 singers as the seen singer test set. To evaluate the model generalization to unseen singers, we prepare 5 utterances from each singer including five males and five females as the additional unseen singer test set. The seen singer test set helps judge if the neural vocoder could synthesize high-fidelity singing voices, and the unseen singer test set helps measure the model's generalization to singer out of domain. At the same time, we take Chinese speech dataset CSMSC for comparison and choose 60 utterances as the test set.

\subsection{Model training}
\quad \textbf{Speaker encoder} We train the speaker encoder with a few large scale multi-speaker datasets following the guidance in~\cite{jia2018transfer}: 1) LibriSpeech Other, which contains 461 hours of speech from a set of 1,166 speakers; 2) VoxCeleb and VoxCeleb2 which contain 139K utterances from 1,211 speakers, and 1.09M utterances from 5,994 speakers, respectively. 3) OpenSinger. We further fine-tune the speaker encoder on the singing voice corpus to ensure it has learned about the distribution of singing voices. Speaker encoder has been trained until convergence (around 50k iterations) as shown in appendix C in the supplementary materials.

\textbf{Multi-Singer} The generator in Multi-Singer takes 80-band mel-spectrograms as input, which are extracted in params (FFT:512, hop size:128, window size:512). At the training stage, a multi-resolution STFT loss would be computed by the sum of three different STFT losses as described in appendix D in the supplementary materials. Multi-Singer is trained for 40k steps with RAdam optimizer to stabilize training. Note that we apply pre-training on the generator for the first 100k steps, after which generator and discriminators are jointly trained. When training previous vocoders such as WaveRNN, MelGAN, and Parallel WaveGAN from scratch, GitHub implementations are used for reproducibility and the configurations follow their original papers.

\subsection{Corpus Verification}
In order to validate the audio quality of OpenSinger, we train several state-of-the-art neural vocoders such as WaveRNN~\cite{kalchbrenner2018efficient}, MelGAN~\cite{kumar2019melgan} and Parallel WaveGAN~\cite{yamamoto2020parallel}. We then conduct evaluations on synthetic seen singer samples\footnote{Audio samples are available at \url{https://Multi-Singer.github.io/}}, and the evaluation matrix has been discussed in appendix B in the supplementary materials.

Our further evaluation lies on the corpus scale. It's well-known that high-fidelity TTS is data-hungry and hence model pre-training is essential in practice. Researchers~\cite{cooper2020pretraining} used to perform the warm-start strategy to overcome data shortage. To explore the effectiveness of additional speech data in SVS training, we pre-train vocoder on a large speech corpus CSMSC (200k steps) and fine-tune the model on OpenSinger until convergence (500k steps). 

Corpus verification results have been introduced in Table~\ref{table:dataset}, and we come to the following conclusions:

\begin{itemize}
  \item OpenSinger simulates the real-world singing voice distribution, and state-of-the-art neural vocoders perform as well as previous papers say. Robust models could be trained using OpenSinger, demonstrating the effectiveness of this dataset.
  \item A slight performance drop appears when using the speech pre-train strategy, which indicates that additional speech data is not needed since OpenSinger is large enough for high-quality singing voice modeling.
\end{itemize}

\begin{table}[H]
  \centering
    \vspace{-2mm}
  \small
  \begin{tabular}{cccc}
  \toprule
   Model            & Dataset                & MOS & FDSD      \\
  \midrule
  WaveRNN          & Singing Voice          & 3.59$\pm$0.15      &0.385      \\
  MelGAN           & Singing Voice          & 3.24$\pm$0.10      &0.864      \\
  Parallel WaveGAN & Singing Voice          & 3.52$\pm$0.12      &0.484      \\
  Parallel WaveGAN & Speech + Singing Voice & 3.49$\pm$0.10      &0.488      \\
  \bottomrule
  \end{tabular}
  \caption{MOS results with 95\% confidence intervals and FDSD for corpus verification.}
    \vspace{-2mm}
\label{table:dataset}
\end{table}
      
\subsection{Multi-band generation}
To verify the effectiveness of the proposed multi-band generator in Multi-Singer, we conduct comparison with competing multi-band architectures such as multiband-MelGAN and multiband-WaveRNN. For fair comparison, we train these models from scratch under the setting in section~\ref{section:Experimental Setup} and implement MOS assessments and real-time factor (RTF) evaluation. From the experimental results in Table~\ref{table:multi-band}, we draw the following conclusions: 1) Due to the auto-regressive architecture, Multi-band WaveRNN achieves the best performance and generates the most natural sounds, which limits overall generation speed on the other hand. 2) Multi-band MelGAN could achieve fast singing voice synthesis, while a distinct quality degradation in audio comes. 3) as for Multi-Singer, because of the multi-band architecture adapted towards characteristic differences among frequency bands, the non-autoregressive generator could adjust singing voice synthesis in both speed and quality

\begin{table}[H]
  \centering
  \small
   \vspace{-2mm}
  \begin{tabular}{ccc}
  \toprule
  Model                                  & MOS  & RTF   \\
  \midrule 
  Multi-band MelGAN                      & 3.21$\pm$0.10 & 0.002 \\
  Multi-band WaveRNN                     & 3.58$\pm$0.13 & 0.350 \\
  Multi-Singer                                 & 3.98$\pm$0.06 & 0.008 \\
  \bottomrule  
\end{tabular}
  \caption{MOS results with 95\% confidence intervals and real-time factor (RTF) of each multi-band generation methods.}
  \vspace{-2mm}
  \label{table:multi-band}
  \end{table}

  \begin{table*}[]
    \centering
    \small
    \vspace{-2mm}
    \begin{tabular}{cccccc}
    \toprule
    \multirow{2}{*}{Model} & \multirow{2}{*}{Train FDSD} & \multicolumn{2}{c}{Seen test singer} & \multicolumn{2}{c}{Unseen test singer} \\
                           &               & MOS        & Cosine similarity   & MOS  & Cosine similarity  \\
    \midrule                       
    WaveRNN (autoregressive) & 0.385      & 3.53$\pm$0.12            & 0.923         & 3.59$\pm$0.15            & 0.931          \\
    SC-WaveRNN (autoregressive) & 0.525   & 3.59$\pm$0.13            & 0.958        & 3.63$\pm$0.13           & 0.970          \\
    MelGAN                 & 0.864        & 3.22$\pm$0.12            & 0.964         & 3.24$\pm$0.10           & 0.943          \\
    Parallel WaveGAN       & 0.484        & 3.51$\pm$0.12            & 0.934        & 3.52$\pm$0.12           & 0.944          \\
    Multi-Singer                 & 0.412        & 3.96$\pm$0.09            & 0.959        & 3.98$\pm$0.06           & 0.967             \\
  
    \bottomrule  
  \end{tabular}
    \caption{MOS results with 95\% confidence intervals, FDSD and cosine similarity for multi-singer modeling. Note that the quality of the added out-of-domain unseen singer recordings is lower than that of seen speaker recordings. Therefore we do not conduct evaluation across seen and unseen singer test set.}
    \vspace{-2mm}
    \label{table:speaker}
    \end{table*}
  
    \begin{table*}[]
      \centering
      \small
      \vspace{-2mm}
      \begin{tabular}{ccccccc}
      \toprule
      \multirow{2}{*}{Model} & \multirow{2}{*}{Train FDSD} & \multirow{2}{*}{RTF} & \multicolumn{2}{c}{Seen test singer} & \multicolumn{2}{c}{Unseen test singer} \\
                             &                 &              & MOS        & Cosine similarity  & MOS  & Cosine similarity   \\
      \midrule   
      w/o MB-generator       & 0.432           & 0.031        & 3.97$\pm$0.07   &0.958        & 4.00$\pm$0.08        &0.966             \\
      w/o SCD                & 0.445           & 0.008        & 3.68$\pm$0.09   &0.952        & 3.70$\pm$0.08        &0.964             \\
      w/o SPL                & 0.461           & 0.008        & 3.81$\pm$0.10   &0.951        & 3.83$\pm$0.09        &0.963            \\
      Baseline (Multi-Singer)      & 0.412           & 0.008        & 3.96$\pm$0.09   &0.959        & 3.98$\pm$0.06        &0.968             \\    
      \bottomrule  
    \end{tabular}
      \caption{MOS results with 95\% confidence intervals, FDSD and cosine similarity for ablation study of each component.}
      \vspace{-4mm}
      \label{table:ablation}
      \end{table*}

\subsection{Multi-singer modeling}

To perform Multi-Singer's better generalization to multiple unseen singers, we evaluate on the synthetic singing voices and compare them with competing architectures. For a fair comparison, we train these vocoders on our proposed multi-singer dataset OpenSinger and use the same experiment setup described in Section~\ref{section:Experimental Setup}. We implement MOS assessments and present objective evaluations such as Fréchet Deep Speech Distances (FDSD), and speaker cosine similarity as shown in Table~\ref{table:speaker}. We have attached the concrete matrix in appendix B in the supplementary materials. Note that the quality of the added out-of-domain unseen singer recordings is lower than that of seen speaker recordings. Therefore we do not conduct meaningless evaluation across seen and unseen singer test set.

Experimental results represent the robustness of Multi-Singer and its outperform capability of unseen singer modeling. We come to conclusions as follows: 1) Architectures such as MelGAN and Parallel WaveGAN haven't explicitly introduced methods for multi-singer adaptation, thus an unavoidable degradation occurs when modeling singing voices of unseen singers. 2) Singer Conditional WaveRNN (SC-WaveRNN) introduces singer-embedding as additional information to control the singer identity during inferences, while the large computational cost slows down generation and increase the difficulty of further application; and 3) as for Multi-Singer, with the prominent capability to perceive singer identity without extra computation during generation, it adjusts singing voice synthesis in both speed and quality.

\subsection{Ablation study}

We conduct ablation studies under the settings in Section~\ref{section:Experimental Setup} to verify the effectiveness of several components in Multi-Singer, including 1) multi-band generator; 2) singer conditional discriminator; and 3) singer perceptual loss. Table~\ref{table:ablation} shows the mean opinion score of audio quality as assessed via human listening tests and objective evaluation results. Our analysis leads to the following conclusions: Replacing \textbf{multi-band generator} with a full band generator causes a significant decrease in generation speed, and we observe that the quality of the synthetic singing voices drops slightly at the same time. The absence of \textbf{Singer Conditional Discriminator (SCD)} results in decreased cosine similarity scores on unseen singers, which suggests that the modified vocoder has difficulties in capturing singer identity. Because abundant singer representations are embedded in the spectrum envelope of singing voices, removing the frequency-domain auxiliary objective \textbf{Singer Perceptual Loss (SPL)} could weaken vocoder in the reconstruction of singer representations. Our ablation study shows that the baseline model Multi-Singer can speed up waveforms generation and efficiently reestablish the singer identity of unseen singers in singing voices.

\subsection{Singing voice synthesis system}
To verify the effectiveness of Multi-Singer in singing voice synthesis systems, we adopt a modified FastSpeech 2 as an acoustic model to convert the words of songs into acoustic features and build an overall system. We have attached the modified FastSpeech 2 in appendix F in the supplementary materials. During training, the configuration follows prior work~\cite{ren2020fastspeech}. Since the F0 and duration are usually known in singing voice synthesis, we remove the pitch and duration prediction, taking the real F0 and phoneme duration as input. FastSpeech 2 converts lyrics, F0, duration and singer embedding into the mel-spectrogram, Multi-Singer converts the mel-spectrogram into singing voices. We perform a MOS scoring test for synthetic audio and get results in Table~\ref{table:system}. Experimental results show that Multi-Singer combined with FastSpeech 2 could generate high-quality singing voices and demonstrate the strong robustness of Multi-Singer in the singing voice synthesis system.

\begin{table}[]
  \centering
  \small
  \begin{tabular}{ccccccc}
  \toprule
  Model                           & MOS \\
  \midrule
  FastSpeech 2 + Multi-Singer           &  3.95$\pm$0.07    \\
  Recording                       &  4.03$\pm$0.09    \\
  \bottomrule
\end{tabular}
\caption{The MOS results with 95\% confidence intervals on each Singing voice synthesis system.}
\vspace{-4mm}
\label{table:system}
\end{table}

\section{conclusion}
We released OpenSinger, a large-scale, multi-singer Chinese singing voice dataset. To our best knowledge, OpenSinger was the first Chinese open dataset towards high-fidelity singing voice synthesis, which we hope would accelerate singing voice synthesis research in the community. To speed up waveforms generation and enhance the capability of vocoder in multi-singer modeling, we proposed Multi-Singer, a fast multi-singer singing voice vocoder. We attached singer conditional discriminator and conditional adversarial training objective to improve singer identity reconstruction.  To assist Multi-Singer sense singer bias between the synthetic and reference singing voices and supervise singer representations reconstruction, we introduced singer perceptual loss as auxiliary loss function. The corpus evaluation demonstrated the effectiveness of OpenSinger for singing voice synthesis researches. Further experimental results showed that Multi-Singer could speed up the generation and synthesize high-fidelity singing voices of unseen singers. Our further experiments proved that Multi-Singer achieved strong robustness in the singing voice synthesis system. For future work, we will continue to study model generalization to different emotions.

\section{Acknowledgements}
This work was supported in part by the National Key R\&D Program of China under Grant No.2020YFC0832505, National Natural Science Foundation of China under Grant No.61836002, No.62072397 and Zhejiang Natural Science Foundation under Grant LR19F020006.


\bibliographystyle{ACM-Reference-Format}
\bibliography{sample-base}

\appendix
\section{Statistics}

As shown in Github pages\footnote{Statistical results are available at \url{https://Multi-Singer.github.io/}}, there come differences between speech data and singing voice data. To be more specific, 1) Singing voices vary from expression and style, including richer emotional information. 2) Singing voices usually have a high sampling rate, which causes a wider spectrogram band in the frequency domain and more high-frequency parts.

\section{Evaluation}
\subsection{Mean Opinion Scores}
All our Mean Opinion Score (MOS) tests are crowdsourced and conducted by native speakers. We refer to the rubric for MOS scores in~\cite{protasio_ribeiro_crowdmos_2011}, and the scoring criteria has been included in Table~\ref{matrix:naturalness} for completeness. The samples are presented and rated one at a time by the testers. 

\begin{table}[H]
  \vspace{-2mm}
  \begin{tabular}{ccc}
  \toprule
  Rating & Naturalness & Definition                           \\
  \midrule
  1      & Bad        &  Very annoying and objectionable dist. \\
  2      & Poor       &  Annoying but not objectionable dist. \\
  3      & Fair       &  Perceptible and slightly annoying dist\\
  4      & Good       & Just perceptible but not annoying dist. \\
  5      & Excellent  & Imperceptible distortions\\
  \bottomrule
  \end{tabular}
  \caption{Ratings that have been used in evaluation of speech naturalness of synthetic and ground truth samples.}
  \vspace{-4mm}
    \label{matrix:naturalness}
\end{table}

\subsection{Fréchet Deep Speech Distances}
Fréchet Deep Speech Distances (FDSD) judges the quality of synthetic audio samples based on their distance to a reference set. These distances are conceptually similar to the FID (Fréchet Inception Distance). The energy distance can be combined with GAN-based learning, further improving on either individual technique. As the paper~\cite{gritsenko2020spectral} says, FDSD is a proper scoring rule with respect to the distribution over spectrograms of the generated waveform audio.

\subsection{Cosine similarity}
Cosine similarity is an objective metric that measures speaker similarity among multi-singer audio. $\operatorname{cos\_sim}(A, B)=\frac{A \cdot B}{|| A|||| B||}$. We also compute the average cosine similarity between embeddings extracted from synthetic speech and the ground truth embeddings to measure the speaker similarity performance objectively. Embeddings of utterances from the same speaker have high cosine similarity, while those from different speakers are far apart in the embedding space.

\section{speaker Encoder}

Speaker verification verifies the speaker identity and tells if the representations come from a related speaker. A speaker-discriminative neural encoder~\cite{wan2018generalized} on a speaker verification (SV) task using a state-of-the-art generalized end-to-end loss. After training on a large amount of data, the speaker encoder could attain robust representations that capture an ample singer identity space. As a result, speaker encoders are usually used for feature extraction, which effectively captures the audio's long-term speaker identity.

Inspired by previous work that has reported the effectiveness of combining a well-trained d-vector model with a TTS model, we build a speaker encoder that projects the mel-spectrogram from the speech utterance to a 256-dimensional. Generalized end-to-end loss (GE2E) makes the training of speaker verification models more efficient than the previous tuple-based end-to-end (TE2E) loss function. During the training process as shown in appendix C in the supplementary materials, the speaker encoder could reduce intra-speaker variance and separate different speakers apart.

\section{Multi-resolution STFT loss details}
Here we introduce details of the multi-resolution STFT loss.
\begin{table}[H]
  \centering
  \begin{tabular}{cccc}
  \toprule
  FFT size & Frame shift & Window size \\
  \midrule
  1024     & 600         & 120        \\
  2048     & 120         & 250         \\
  512      & 240         & 50         \\
 \bottomrule
 \end{tabular}
\caption{The details of the multi-resolution STFT loss. A hanning window was applied before the FFT process.}
\vspace{-2mm}
\label{table:stftloss}
\end{table}

\section{WaveNet Block}
The architecture of WaveNet block~\cite{oord2016wavenet} in generator has been shown in Figure~\ref{fig:appendix_arch}. X and Mel denote noise and mel-spectrogram, respectively. Noise passes through the dilated convolution layers, and X, Mel are divided into xa, xb and sa, sb, respectively. After the sigmoid-tanh calculation, the processed feature passes through two fully-connected networks and output H and X, which would be fed into the next dilated convolution layer.

\section{Modified FastSpeech 2}
\begin{figure}[]
  \centering
  \vspace{-2mm}
  \includegraphics[width=0.4\textwidth]{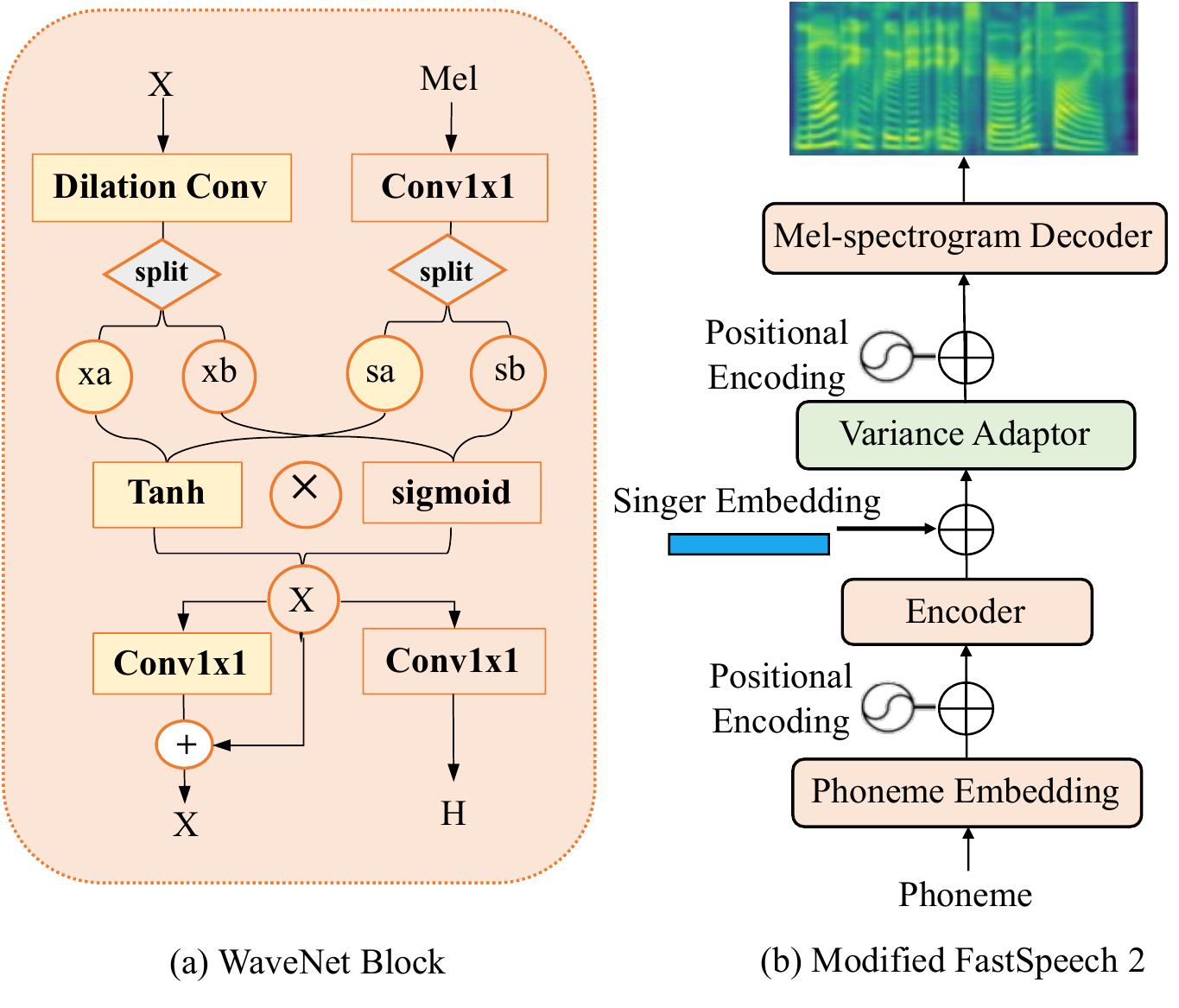}
  \caption{Architectures.}
  \label{fig:appendix_arch}
  \vspace{-3mm}
  \end{figure}

As shown in Figure~\ref{fig:appendix_arch}, our modified FastSpeech 2 converts lyrics, F0, duration, and singer embedding into the mel-spectrogram, after which Multi-Singer converts the mel-spectrogram into singing voices.

\end{document}